\documentclass[aps,pre,twocolumn]{revtex4-1}
\usepackage{amsmath,bm,epsfig,,subfigure}
\usepackage{color}
\usepackage[normalem]{ulem}

\def\Fbox#1{\vskip1ex\hbox to 8.5cm{\hfil\fboxsep0.3cm\fbox{%
  \parbox{8.0cm}{#1}}\hfil}\vskip1ex\noindent}  
\newcommand{\Hes}{\B{\C H}}

\newcommand{\B}[1]{{\bm{#1}}}
\newcommand{\C}[1]{{\mathcal{#1}}}    
\let \= \equiv \let\*\cdot \let\~\widetilde \let\-\overline

\begin{document}
\title{Atomistic Simulations of Magnetic Amorphous Solids: Magnetostriction,\\
Barkhausen noise and novel singularities}
\author{Ratul Dasgupta, H. George E. Hentschel, Itamar Procaccia and Bhaskar Sen Gupta}
\begin{abstract}
We present results of atomistic simulations of a new model of a magnetic amorphous solid subjected to external mechanical strains and magnetic fields. The model employed offers new perspectives on important effects like Barkhausen noise and magnetostriction. It is shown that
the plastic response in such systems exhibit singularities characterized by unexpected exponents requiring careful theoretical reasoning.  The spatial structure of the plastic events requires a new coarse grained elasto-magnetic theory which is provided here.
\end{abstract}
\maketitle

Our understanding of the mechanical properties of amorphous solids has received a strong boost from atomistic simulations, leading to novel theories of
plastic behavior in such systems. In particular localized plastic events could be understood as the appearance of effective Eshelby inclusions in the elastic matrix \cite{99ML,04ML}. This relation allows analytic derivations, culminating recently with a theory of shear localization in amorphous solids, leading to shear bands and material failure \cite{12DHP,13JGPS}. As long as only mechanical properties are involved, the nature of the plastic singularities is now fully understood, being dominated by a sequence of simple saddle node bifurcations as the material is strained. This leads to a high degree of universality in the nature of plastic events in widely different amorphous solids ranging from Lennard-Jones binary mixtures to metallic glasses \cite{12DKP}.

In contrast, atomistic simulations of externally strained magnetic amorphous solids in the presence of magnetic fields are at an earlier stage of development.  Theoretical considerations indicate however that new singularities and richer physics are expected from the presence of two independent control parameters, mechanical and magnetic
\cite{12HIP}. In this Letter we report on such simulations that indeed appear to present much new physics. Here we discuss results on athermal quasi-static (AQS) simulations in two-dimensions. Extensions to 3D and to finite temperature and finite strain rates will be reported elsewhere.

{\bf The model}: The model we employ is in the spirit of the Harris, Plischke and Zuckerman (HPZ) Hamiltonian \cite{73HPZ}
but with a number of important modifications to conform with the physics of amorphous magnetic solids \cite{12HIP}. We write the Hamiltonian as
\begin{equation}
\label{umech}
U(\{\B r_i\},\{\B S_i\}) = U_{\rm mech}(\{\B r_i\}) + U_{\rm mag}(\{\B r_i\},\{\B S_i\})\ ,
\end{equation}
where $\{\B r_i\}_{i=1}^N$ are the 2-D positions of $N$ particles in an area $L^2$ and $\B S_i$ are spin variables. The mechanical part $U_{\rm mech}$ is chosen to represent a glassy material with a binary mixture of 65\% particles A and 35\% particles B,
with Kob-Anderson Lennard-Jones potentials having a minimum at positions $\sigma_{AA}=1.17557$, $\sigma_{AB}=1.0$ and $\sigma_{BB}=0.618034$ for
the corresponding interacting particles. These values are chosen to guarantee good glass formation and avoidance of crystallization. The energy parameters chosen are $\epsilon_{AA}=\epsilon_{BB}=0.5$
$\epsilon_{AB}=1.0$ in units for which the Boltzmann constant equals unity. All the potentials are truncated at distance 2.5$\sigma$ with two continuous derivatives. Particles A carry spins $\B S_i$; the B particles are not magnetic.

The magnetic potential needs to be modeled to best fit a particular material, and different materials will have somewhat different magnetic interactions. For concreteness we consider here the spins $\B S_i$ to be
classical $xy$ spins; the orientation of each spin is then given by an angle $\phi_i$. We also denote by $\theta_i({\bf r}_i)$ the local preferred easy axis of anisotropy, and end up with the magnetic contribution to the potential energy in the form \cite{12HIP}:
\begin{eqnarray}
&&U_{\rm mag}(\{\B r_i\}, \{\B S_i\}) = - \sum_{<ij>}J(r_{ij}) \cos{(\phi_i-\phi_j)}\nonumber\\&&-  \sum_i K_i\cos{(\phi_i-\theta_i(\{\B r_i\}))}^2-  \mu_A B \sum_i \cos{(\phi_i)} \ .
\label{magU}
\end{eqnarray}
Here $r_{ij}\equiv |\B r_i-\B r_j|$ and the sums are only over the A particles. For a discussion of the physical significance of each term the reader is referred to Ref.~\cite{12HIP}. It is important however to stress that in our model (in contradistinction with the HPZ Hamiltonian \cite{73HPZ} and also with the Random Field Ising Model \cite{93Seth}), the exchange parameter $J(\B r_{ij})$ is a function of a changing inter-particle position (either due
to external strain or due to non affine displacements, and see below). We choose the monotonically decreasing form
$J(x) =J_0 f(x)$ where $f(x) \equiv \exp(-x^2/0.28)+H_0+H_2 x^2+H_4 x^4 $ with
$H_0=-5.51\times 10^{-8}\ ,H_2=1.68 \times 10^{-8}\ , H_4=-1.29 \times 10^{-9}$.
This choice cuts off $J(x)$ at $x=2.5$ with two smooth derivatives.  In our case $J_0=1$.

Another important difference is that in our case
the local axis of anisotropy $\theta_i$ is {\em not} random, but is determined by the local structure: define  the matrix $\B T_i$:
\begin{equation}
T_i^{\alpha\beta} \equiv \sum_j J( r_{ij})  r_{ij}^\alpha r_{ij}^\beta/\sum_j J( r_{ij}) \ .
\end{equation}
The matrix $\B T_i$ has two eigenvalues in 2-dimensions that we denote as $\kappa_{i,1}$ and $\kappa_{i,2}$, $\kappa_{i,1}\ge \kappa_{i,2}$. The eigenvector that belongs to the larger eigenvalue $\kappa_{i,1}$ is denoted by $\hat {\B n}$. The easy axis of anisotropy is given by by $\theta_i\equiv \sin^{-1} (|\hat n_y|)$. Finally the coefficient $K_i$ which now changes from particle to particle is defined as
\begin{equation}
K_i \equiv \frac{5[\sum_j J( r_{ij})]^2 (\kappa_{i,1}-\kappa_{i,2})^2} {J_0\sigma^4_{AB}} \ .
\end{equation}
This definition guarantees both that $K_i$ has units of energy and that the contribution due to anisotropy will vanish when the local neighborhood of
the $i$th particles is isotropic. This choice of $K_i$ is not unique but rather represents the essential physics
of local anisotropy. The last term in Eq. (\ref{magU}) is
the interaction with the external field $B$. We have chosen $\mu_A$ in the range [-0.08,0.08]. At the two extreme values all the spins are aligned along the direction of $\B B$.

\begin{figure}
\includegraphics[scale = 0.35]{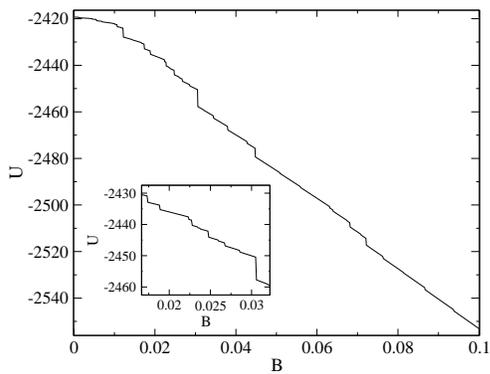}
\caption{The change of energy during a quasi-static ramping of the magnetic field $\B B$. The inset shows that the sharp drops which are plastic (irreversible) events continue to small scales.}
\label{UvsB}
\end{figure}

\begin{figure}
\vskip 1 cm
\includegraphics[scale = 0.27]{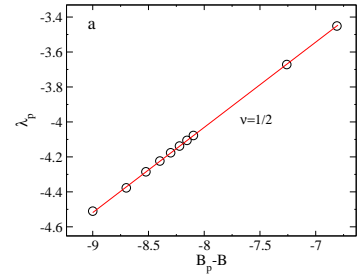}
\vskip 0.8 cm
\includegraphics[scale = 0.27]{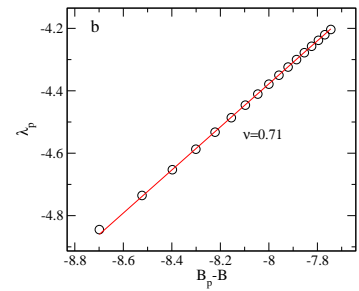}
\vskip 0.8 cm
\includegraphics[scale = 0.27]{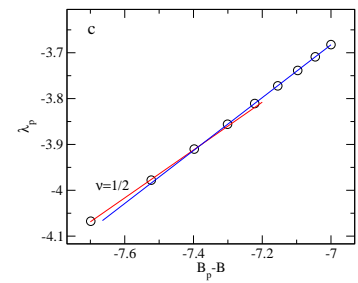}
\caption{Scaling laws and apparent scaling laws for $\lambda$ vs $B_p-B$ for different values of $B_p$. For $B_p$ very
small or very large (panel a is for large $B$) we expect and find a slope 1/2. There exists a value of $B_p=B_c$ where we expec a slope of 3/4, cf. panel b for $B_p$ close to $B_c$. Finally, for any other $B_p$ we expect and find a cross over to a slope 1/2
very near the instability.}
\label{lamvsB}
\end{figure}

{\bf New exponents}: to initiate the study of the interesting physics exhibited by this model we show in Fig. \ref{UvsB} how the energy
changes when the magnetic field increases without external strain. The increase of $\B B$ is done quasi-statically, performing energy minimization after every step of increase of $\B B$. Reversible smooth changes in the energy are
punctated by (irreversible) drops in energy which are caused by localized magneto-plastic events.
As usual, these events are associated with an eigenvalue of the Hessian matrix going to zero, while at the same
time the associated eigenfunction that is delocalized far from the instability gets localized on $n$ particles where
$n$ can be much smaller and independent of $N$ or it may scale like $N^\alpha$ depending on the value of $B$ \cite{99PDS}. As stated above, in amorphous solids subjected to pure mechanical strains
the eigenvalue goes to zero with a characteristic $\nu=1/2$ exponent, i.e. $\lambda \sim (\gamma_p-\gamma)^{1/2}$ where $\gamma$ is
the magnitude of the external strain and $\gamma_p$ its value at the occurrence of the plastic event \cite{12DKP}. Our first interesting
finding is that this is no longer the case here, and $\nu$ can differ from 1/2.
In Fig. \ref{lamvsB} we show log-log plots of $\lambda$ vs. $B_p-B$ close to an instability at $B_p$, indicating a complex critical behavior $\lambda\sim (B_p-B)^\nu$ with exponents $\nu$ that can differ from 1/2.
For $B_p$ very small and very large (see for example panel a) the slope in the log-log plot is very close to 1/2.
There is a value of $B_p$, which we denote as $B_c$ for which we expect a slope of 3/4. In panel b $B_p$ is in the vicinity of $B_c$ and $\nu\approx 0.71$. For $0<B_p<B_c$
and for $B_p>B_c$ we expect a crossover between a slope higher than 1/2 to a slope 1/2 in the vicinity of $B_p$.

\begin{figure}
\includegraphics[scale = 0.30]{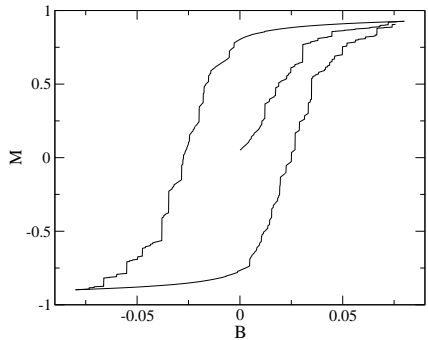}
\caption{Hysteresis loop of the magnetization as a function of magnetic field, including the increase from the freshly quenched glass, then decreasing the field and increasing it again.}
\label{Bark}
\end{figure}

To understand these results we recall that the eigenvalue that goes soft belongs to the
Hessian matrix $\Hes$ which in the present case takes on the form \cite{12HIP}:
\begin{equation}
\label{Hesa}
\Hes =
\begin{pmatrix}
  \frac{\partial^2U}{\partial \B r_i\partial \B r_j} & \frac{\partial^2U}{\partial \B r_i\partial \phi_j} \\
  \frac{\partial^2U}{\partial \phi_i\partial \B r_i}  & \frac{\partial^2U}{\partial \phi_i\partial \phi_j}
\end{pmatrix}
\end{equation}
The eigenfunctions associated with the eigenvalues $\lambda_k$ are denoted $\B \Psi^{k}$. In terms of these objects
we have an exact result for the change of $\lambda$ with $B$ \cite{12HIP}:
\begin{equation}
\frac{\partial \lambda_k}{\partial B}{\bf |}_{\gamma}   =   c^{(b)}_{kk} - \sum_\ell \frac{a^{(b)}_\ell [b^{(r)}_{kk\ell}+b^{(\phi )}_{kk\ell}]}{\lambda_\ell} .
\label{diff}
\end{equation}
The precise definition of all the coefficients is given explicitly in Ref. \cite{12HIP}. For the present purpose it is
sufficient to know that $c_{kk}^{(b)}$, and both $b_{kk\ell}$ cannot be zero or singular, but on the other
hand $a_\ell^{(b)}$ may vanish at some value of $B$. This coefficient is defined as
\begin{equation}
a_\ell^{(b)} \equiv \B \Xi^{(b)}\cdot \B \Psi^{\ell}\ ,  ~\Xi^{(b)} = \begin{pmatrix}
  0\\
  \B \Xi^{(b, \phi )}
\end{pmatrix} \ ,  ~ \B \Xi^{(b, \phi )}\equiv   \frac{\partial^2U}{\partial B \partial \phi} .
\label{defxi}
\end{equation}
Clearly, when one of the eigenvalues $\lambda_p$ goes to zero, Eq. (\ref{diff}) simplifies to one dominant term
\begin{equation}
\frac{\partial \lambda_p}{\partial B}{\bf |}_{\gamma}   \approx   -  \frac{a^{(b)}_p C }{\lambda_p} .
\label{diff1}
\end{equation}
Where $C$ is a constant when $B\to B_p$. We have checked that in our system there exists a gap to the next eigenvalue justifying the last equation. Examining our Hamiltonian we discover that
\begin{equation}
a_p^{(b)}=\sum_{i=1}^N\sin\phi_i \Psi_i^{p} \ .
\end{equation}
Noting that $\B \Psi^{(p)}$ gets localized on $n\ll N$ particles and that it is normalized, we expect
\begin{equation}
a_p^{(b)} \approx \frac{1}{\sqrt{n}}\sum_{i=1}^n \sin \phi_i =\sqrt{n}\langle \sin \phi \rangle_n \ ,
\end{equation}
where the notation $\langle \dots \rangle_n$ means an average over the particles on which the eigenfunction is localized. From this result we conclude that whenever $\langle \sin \phi \rangle_n$ is not zero, Eq. (\ref{diff1}) will lead
to exponent $\nu=1/2$. Indeed for small $B$ the spins point out in the quasi-random local anisotropy axis, and
$\langle \sin \phi \rangle_n\ne 0$. Also when $B$ is very large, most of the spins point out in the $x$ direction,
but {\em unstable} modes will consist of spins pointing otherwise, so that again $\langle \sin \phi \rangle_n\ne 0$.
But for an intermediate value of $B$, where the drops in $U$ in Fig. \ref{UvsB} or in the magnetization in Fig.
\ref{Bark} are largest, the most unstable mode will consist of spins pointing almost opposite to the applied field, with
$\phi_i \approx \pi$. We thus define the value of $B=B_c$ as that point for which (in the thermodynamic limit) $\langle \sin \phi \rangle_n=0$. For $B_p$ in the vicinity of $B_c$ we write $\phi_i=\pi+\Delta \phi_i$, and therefore
\begin{equation}
a_p^{(b)} \approx -\frac{1}{\sqrt{n}}\sum_{i=1}^n \sin \Delta \phi_i \approx -\frac{1}{\sqrt{n}}\sum_{i=1}^n \Delta \phi_i
\approx \sqrt {\langle (\Delta \phi_i)^2\rangle_n} \ ,
\end{equation}
by the central limit theorem. Since $\langle (\Delta \phi_i)^2\rangle_n=0$ at $B-B_c$ we expect in the vicinity of
$B_c$ to have $\langle (\Delta \phi_i)^2\rangle_n \propto (B_c-B)$. Substituting this in Eq. (\ref{diff1}) we get
immediately $\nu=3/4$ as is seen in the numerics. Note that this result is only relevant for $B_c$. At any other
value of $B$ we expect to see $\nu=1/2$ or a cross over to this value for $B\to B_p$.

\begin{figure}
\vskip 0.5cm
\includegraphics[scale = 0.33]{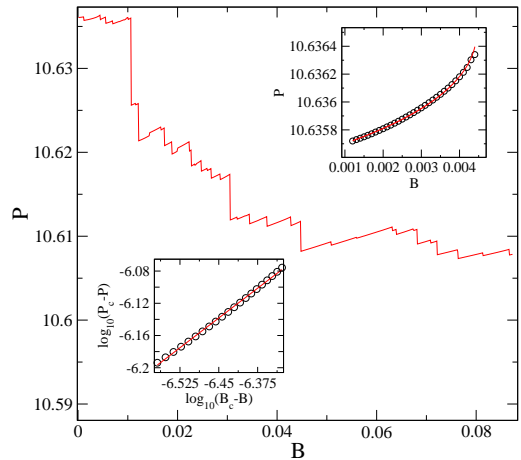}
\caption{Pressure as a function of applied magnetic field. The magnetosriction coefficient is positive in the present case.
Upper inset: the fit of the data to Eq. (\ref{predict}). Lower inset: a log log plot of the pressure vs. $B-B_p$ very near
the plastic drop with a slope 1/2.}
\label{striction}
\end{figure}

{\bf Magnetostricton}: another interesting physical effect that warrants much more analysis is magnetostriction. This effect can be studied in an NPT ensemble as the change in the volume with changing the magnetic field or in an NVT ensemble by the change in the pressure. We chose here the latter and in Fig. \ref{striction} we display
the pressure $P$ as a function of $B$ in a quasi-static ramp of $B$.
During smooth sections the magnetostriction coefficient is positive in our case (in the NVT ensemble the pressure increases with B) but
the pressure is again punctuated by plastic drops which lead to an overall decrease in pressure. Between the drops the pressure increases
with a tendency towards singularity immediately preceding the drop. To understand the physics displayed here we recall that
the pressure can be written in our system (at $T=0$) as
\begin{equation}
P = \frac{1}{2V}\sum_{i\ne j}\sum_{i} \B f_{ij}\cdot \B r_{ij}\ , \quad \B r_{ij} =\B r_i-\B r_j \ ,
\end{equation}
where $\B f_{ij}$ is the force exerted by the $j$th particle on the $i$th particle. Taking the derivative with respect to
$B$,
\begin{equation}
2V \frac{\partial P}{\partial B} = \sum_{i\ne j}\sum_{i}\frac{\partial \B f_{ij}}{\partial B}\cdot \B r_{ij}+ \sum_{i\ne j}\sum_{i} \B f_{ij}\cdot \frac{\partial \B r_{ij}}{\partial B} \ .
\end{equation}
The first term on the RHS is not expected to be singular when $B\to B_p$, since the derivative there is of the type of $\B \Xi$
of Eq. (\ref{defxi}) which is never singular. The second term is singular however at a plastic event, being a non-affine
coordinate change, proportional to $\Hes^{-1}\cdot \B \Xi$. Near the plastic event this second term is proportional
to the RHS of Eq.~(\ref{diff1}), so that for $B$ small or large we expect
\begin{equation}
P\approx P_p -C_1 (B_p-B) -C_2 \sqrt {B_p-B} \ ,
\label{predict}
\end{equation}
where $P_p$ is the value of the pressure before the plastic drop, and $C_1$ and $C_2$ positive. A fit to this
formula for $B_p\approx 0.0044$ is shown in the insets in Fig. \ref{striction}, explaining the shape of the smooth parts {\em and} the plastic punctuations in Fig. \ref{striction}. For $B_p$ close to $B_c$ the new exponent $\nu=3/4$
will change Eq. (\ref{predict}) in an obvious way.

{\bf Barkhausen noise}: needless to say, our model allows for a detailed study of the Barkhausen noise \cite{10DMW}. In Fig. \ref{Bark} we exhibit the magnetization as a function of $B$. The shorter line represents the change in magnetization as $B$ is increased
starting from the freshly quenched glass. Upon saturation, the magnetic field is inverted until the magnetization
become -1, where the field is again increased to display the well known hysteresis loop.
The statistics of Barkhausen noise was studied extensively in the context of the Random Field Ising Model with
strong claims for universality \cite{95PDS}. The model proposed here promises excellent testing grounds of such claims within
a model of actual amorphous solids with magnetic properties.
\begin{figure}
\includegraphics[scale = 0.27]{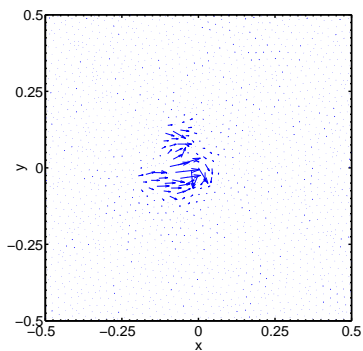}
\caption{A typical non-affine magnetic displacement field; shown are $\Delta \B S_i$ during the event.}
\label{nonaff}
\end{figure}

{\bf Coarse grained model}: Finally we should comment on the interesting subject of the non-affine displacements in this model and how to
provide a theory for them. In amorphous solids undergoing purely mechanical loading the plastic events are well
described by the Eshelby theory, being quadrupolar in 2D with obvious generalization to 3D. In the present case
the {\em mechanical} part of the non-affine displacements are still quadrupolar as expected from the Eshelby theory,
but the magnetic non-affine changes are totally different, and may display magnetic topological singularities. An example
of a typical event that displays $\Delta \B S_i$ in a plastic event is shown in Fig. \ref{nonaff}.
To create the analog of the Eshelby theory for those we need first to generalize the Lam\'e-Navier equations to the
present case. A convenient starting point is the free energy functional
\begin{eqnarray}
&&F =\int dx dy \Big[ \mu (u_{ik}-\frac{1}{2} \delta_{ik} u_{\ell\ell})^2 +\frac{1}{2}\kappa u_{\ell\ell}^2-K (m_i n_i)^2\nonumber\\
&&-b \left(u_{ik} -\frac{u_{\ell\ell}}{2}\delta_{ik}\right) m_i m_k +\frac{a}{2} \left(\frac{\partial m_i}{\partial x_k}\right)^2 -B ~m_x\Big] \ ,
\end{eqnarray}
where $\mu$ and $\kappa$ are the shear and bulk moduli and $b$ is the anisotropic magnetoelastic coupling term. $\B n$ is
a random vector field representing the local anisotropy.  The
coupling constant $a$ represents the exchange interaction. For an $xy$ model $m_x=\cos\theta$ and $m_y=\sin \theta$.
Substituting and following the Euler-Lagrange procedure one ends up with the generalized Lam\'e-Navier equations
\begin{widetext}
\begin{eqnarray}
&&\mu\Big (\frac{\partial^2u_x}{\partial x^2}+\frac{\partial^2u_x}{\partial y^2}\Big)+\kappa \frac{\partial}{\partial x} \Big(\frac{\partial u_x}{\partial x}+\frac{\partial u_x}{\partial y}\Big) =k\Big[\cos 2\theta\Big(\frac{\partial \theta}{\partial y}\Big) -\sin 2\theta \Big(\frac{\partial \theta}{\partial x}\Big)\Big]\\
&&\mu\Big (\frac{\partial^2u_y}{\partial x^2}+\frac{\partial^2u_y}{\partial y^2}\Big)+\kappa \frac{\partial}{\partial y} \Big(\frac{\partial u_x}{\partial x}+\frac{\partial u_y}{\partial y}\Big) =k\Big[\cos 2\theta\Big(\frac{\partial \theta}{\partial x}\Big) +\sin 2\theta \Big(\frac{\partial \theta}{\partial y}\Big)\Big]\\
&&a \Big(\frac{\partial^2 \theta}{\partial x^2}+\frac{\partial^2 \theta}{\partial y^2}\Big) = k \Big[ \Big(\frac{\partial u_x}{\partial x} -\frac{\partial u_y}{\partial y}\Big)\sin 2\theta - \Big(\frac{\partial u_x}{\partial y} +\frac{\partial u_y}{\partial x}\Big)\cos 2\theta\Big] +B~\sin\theta +K \sin 2(\theta-\eta(x,y)) \ ,
\end{eqnarray}
\end{widetext}
With $\eta(x,y)$ being the angle of the local easy axis that varies randomly. The first two equations simplify to the standard Lam\'e-Navier equations when $k=0$.
The geometric characters of the nonaffine plastic events are determined by these equations.

In summary, we have offered a new model for an amorphous solid with magnetic interactions. The results of this Letter
indicate a number of interesting aspects of this model that are extremely worthwhile and promise new physics. It is our
aim to elucidate all these aspects in full detail in forthcoming publications.

\end{document}